# RECOGNIZING INDIVIDUALS AND THEIR EMOTIONS USING PLANTS AS BIO-SENSORS THROUGH ELECTRO-STATIC DISCHARGE

Buenyamin Oezkaya
Peter Gloor
MIT Center for Collective Intelligence
pgloor@mit.edu

## ABSTRACT

By measuring the electrostatic discharge of human bodies together with Mimosa Pudica and other plants in response to the human movement we have been able to recognize (a) individuals based on their distinctive pattern of body movements with 66% accuracy as well as (b) positive or negative mood based on their gait characteristics with 85% accuracy. We use the Plant SpikerBox, a device that measures the electrical action potential while also measuring the electrostatic discharge between the electrode on the leaves of a plant and the capacitively coupled human body.

## 1. INTRODUCTION

In this paper, we describe how plants can be integrated as biological parts into an electric field detection unit to sense human behavior and emotions. Embedding plants into sensors is a natural and powerful extension of the Internet of Things. Plants naturally pervade our environment and are significantly less expensive in terms of production, operation, and maintenance costs than any artificial device (Manzella et al. 2013). Furthermore, plants do not fall under privacy and GDPR[1] concerns. Moreover, conventional wireless systems usually store data or transmit data through Bluetooth. Unlike wearables, direct contact between the electrode and the human body is not necessary. Our approach is affordable, wear-free, and fit for long-time monitoring during daily activities. Furthermore, no devices need to be placed on the subject's body, which may have been perceived as uncomfortable or obtrusive. In earlier, related work, researchers have used plants as sensors for environmental monitoring, e.g., pollution, and fires, in agriculture, e.g., for monitoring irrigation, plant health, and the use of chemicals, and area monitoring, e.g., for avalanches, and flooding (Chatterjee et al. 2015). To the best of our knowledge, this use case applying electric field sensing to plants has not been described before.

## 2. MEASURING THE ELECTROSTATIC DISCHARGE OF HUMANS WITH PLANTS

For our experiments, we are using the commercially available Plant SpikerBox[2], an Arduino-based data acquisition system with filters for plant signal measurement with a low bias-current amplifier already installed. For checking the environmental conditions, we used a RaspberryPi-based monitoring system. Using different sensors for soil-moisture, $CO_2$-concentration, temperature, and humidity, we were able to check the environmental conditions change during a measuring session. The changes in all measured fields were negligible.

Human gait has, in the past, been measured through electrostatic discharge (Chen et al. 2012). When a human walks, steps, or jumps, static electricity is produced as a result of friction between the body and clothing. The separation between the human foot and the ground during walking also charges the human body (Li et al. 2018).

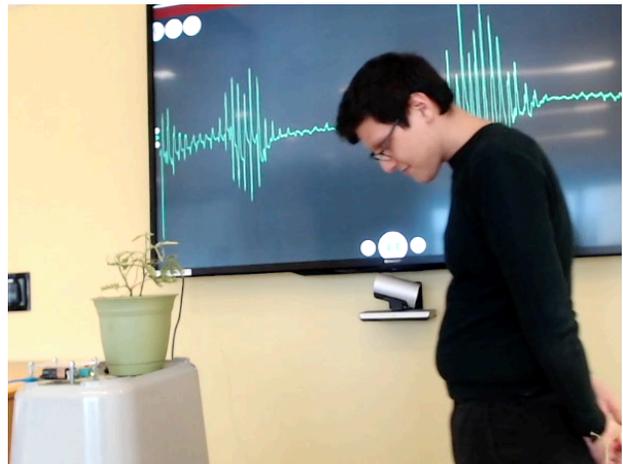

(a)

---

[1] General Data Protection Regulation

[2] https://backyardbrains.com/products/plantspikerbox

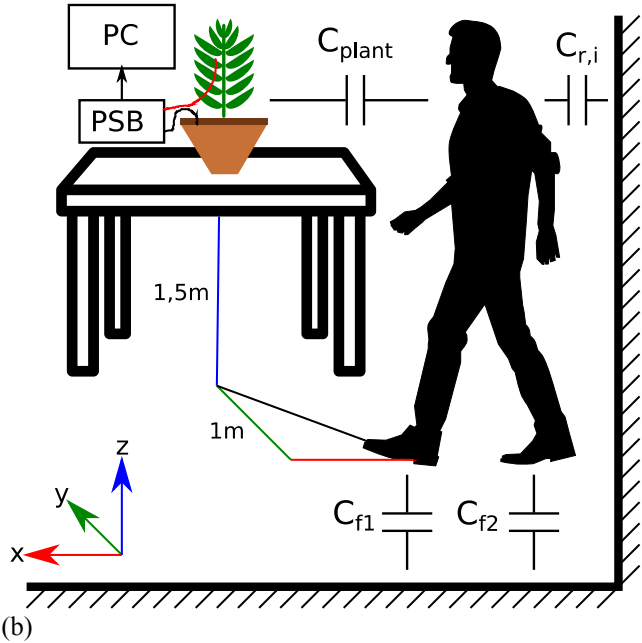

(b)

**Figure 1.** Electrostatic discharge is shown while experimenter walks by Mimosa Pudica which is connected to plant spiker box (PSB) (a) Schematic Diagram of Experiment (b)

This change in static electricity arises when the foot makes or breaks contact with the floor. A negative charge occurs when the foot contacts the ground, while a positive charge is generated when the foot detaches from the ground (Chen et al. 2012). Therefore, it is possible to distinguish a foot making and breaking contact. These effects are known as contact electricity charges. During the walking motion, human capacitance (CB) can be expressed by Equation (1) (Chen et al. 2012). The capacitance between the foot and the Ground (CF) depends on the height of the sole. Both CF and the capacitance relative to nearby objects inside the room (Cr, i) is relative to the position of the individual.

$$\frac{1}{C_B} = \frac{1}{C_{f1}} + \frac{1}{C_{f2}} + \sum_{i=1}^{\infty} \frac{1}{C_{r,i}} = f(t) \qquad (1)$$

The induced current i(t) due to the human walking motion is illustrated in equation (2) and (3). $Q_e$ is the charge induced on the plant electrode. The capacitance between the human body and the plant electrode is $C_{plant}$. S represents the equivalent area between the human body and the plant. X (t) and y(t) illustrate the horizontal distance and vertical distance, thus considering the radial distance to the plant. The permittivity of the air is described with $\varepsilon$. I(t) takes to account the induced current generated by the feet motion without considering the variation of capacitance between the human body and the plant electrode, e.g., stepping in place.

$$i(t) = \frac{dQ_e}{dt} = \frac{\varepsilon S}{\sqrt{x(t)^2 + y(t)^2}} I(t) \qquad (2)$$

$$I(t) \propto C_{plant} \frac{d}{dt}\left(\frac{1}{C_B}\right) \qquad (3)$$

The influence on the amplitude of measured electric field strength caused by environmental factors such as temperature, humidity, and clothing should be taken into consideration (Chubb 2008) as they will cause noticeable variations on amplitudes of measured signals. However, these environmental conditions will not change during a measuring session. The influence of the session itself is discussed in section 4.

**2.1 Experimental Setup**

A high-impedance voltage amplifier that is attached to the plant and capacitively coupled to the human body can sense this change in static electricity. Figure 2 visualizes the key components of our experimental setup. In our approach, we use a single-ended measurement of the plant stem using the soil as the local ground. The electrode can be used for electric field sensing due to the high resistance of the plant. In addition to electric field sensing, we can measure the action potentials of the plant generated by physical stress, like touching the leaves.

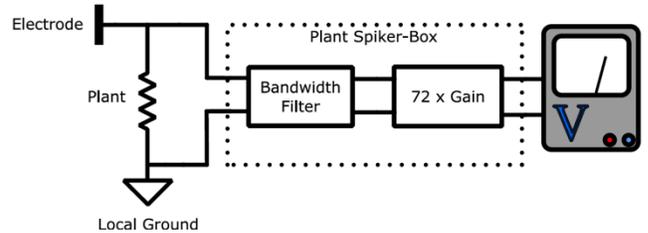

**Figure 2**. Schematic of our Data-Acquisition system

The electrical signals must be converted to features suitable for machine learning. Spectrograms are commonly used for ECG/EEGs (Colomer et al. 2016) and are utilized for audio classification in deep learning applications (Hershey et al. 2017). In the past, researchers have extracted statistical features like MFCCs (Mel-frequency cepstral coefficients) for audio analysis (Tahon & Devillers 2016). The gathered signals in our approach look similar in envelope, shape, and frequency to signals in audio analysis. MFCCs are a small set of 10-20 features that describe the overall shape of a spectral envelope. In this paper, we used the parameters shown in Table 1 for feature extraction using MFCC. The sampling rate is given through the Plant SpikerBox. One reason to calculate MFCCs is to reduce the dimensionality of the amplitude spectrum, as well as to capture its envelope.

| Features : MFCC | |
|---|---|
| Sampling Rate | 10000 |
| Number of MFCCs | 20 |
| Window Size | 2500 |
| Hop Length | 1250 |
| Exponent of Magnitude | 2 |

**Table 1.** Parameters to extract MFCC from a sample

The first step is to measure the energy in the filter banks. Those triangular filters are spaced over the Mel scale (Marechal et al. 2019). The denser resolution in low frequencies compared to higher frequencies fits our approach since most of the information in the electrical signal we are measuring is at lower frequencies. The importance of the envelope in our application of gait-based electrostatic discharge was shown in (Chen et al. 2012). The Window Size and Hop Length for calculating the MFCCs were chosen to fit best for our application of electric field sensing and analyzing human gait of 1-2Hz (Li et al. 2019). Every feature has an additional time dimension to take walking time into account, resulting in additional calculations every 100ms.

**2.2 Applying Machine Learning for Individual Prediction**

Figure 3 describes the process we are applying for analysis, converting the electrical signals captured with the Plant SpikerBox into 20 MFCC signals, which are used to train a machine learning model. For plants, we have used Mimosa Pudica, common basil, thyme, and orchids. We have also been able to obtain the same electrical signal when connecting the electrodes of the plant spiker box to a fresh twig of hazel, beech, or apple tree with the grounding wire connected to the water in the vase where the twig was standing.

Each gathered electrical signal had been trimmed so that every sample had the same length. Subsequently, each gait-sequence was standardized using the Z-Transformation. In doing that, we eliminated the effect of different amplitude dimensions on subsequent analyses. By using MFCCs as our main feature, the dataset was reasonably large enough to use a decision tree based random forest classifier. Good accuracy has been achieved using decision trees to predict dangerous chemicals from electrical signals in plants (Chatterjee et al. 2017). For predicting human gait characteristics using electric field sensing, decision trees have also been shown viable (Li et al. 2019).

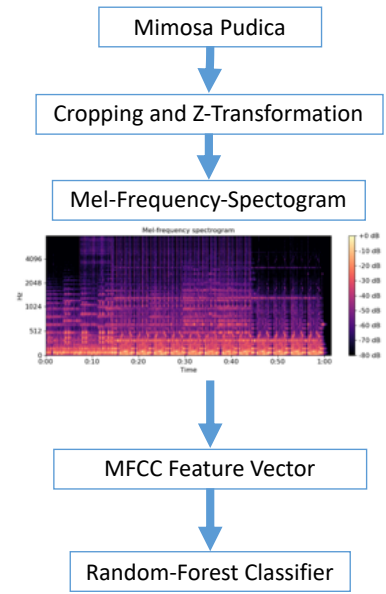

**Figure 3.** The machine learning process for people/emotion recognition

In general, decision trees tend to be robust since they can work with different types of data and outperform other methods such as linear or logistic regression. They have no problem in the case of heterogeneous features or mixing completely different ranges of values (Strobl et al. 2007). To avoid overfitting and to make full use of the experimental data, a 10-fold Cross-validation algorithm was used for the train/test data split. A stratified algorithm was used to ensure that each fold has the same proportion of observations with a given categorical value. An additional shuffling of the data before the stratified 10-fold split was introduced to avoid overfitting even further. For the hyperparameters of our random forest, we used a randomized search optimization using over 7920 possible combinations and 100 iterations (Bergstra & Bengio 2012).

**3. PREDICTING THE MOOD OF DIFFERENT PEOPLE BASED ON THEIR WALKING PATTERN**

Dysphoric mood can be observed in gait patterns of individuals. Characteristics associated with being sad are smaller amplitudes in vertical movements of the upper body and reduced walking speed (Michalak et al. 2009). For our experiment, four individuals have walked "sad" and "happy" accordingly, collecting over 139 samples. Figure 5 shows two samples of z-transformed signals. Signal A shows the induced electric current of someone walking sadly towards a plant over time. The slower footsteps are shown as the spikes appear in much lower frequencies. The increasing amplitude results from the decreasing distance of the capacitively coupled human and the plant. Signal B shows the signal of a human walking happily away from a plant. The spikes are not only higher in amplitude but also occur at a higher frequency. Signal B is shorter than signal A since the walking speed of a happy person is higher. Using the random forest classifier, an accuracy of 85.5 % was achieved (Cohens κ=0.711, AUROC=0.856). Table 2 shows an overview of the used hyperparameters for the "sad" and "happy" classification.

| Classifier : Random-Forest | |
|---|---|
| N | 139 |
| Number of Classes | 2 |
| Cross-validation | 10-Fold CV and 80/20 Train/Test Split |
| Estimators | 100 |
| Minimum Samples to split Node | 5 |
| Minimum Samples per Leaf | 4 |
| Maximum depth of Tree | 100 |
| Samples drawn with replacement | False |

**Table 2**. Hyperparameters of our random forest classifier

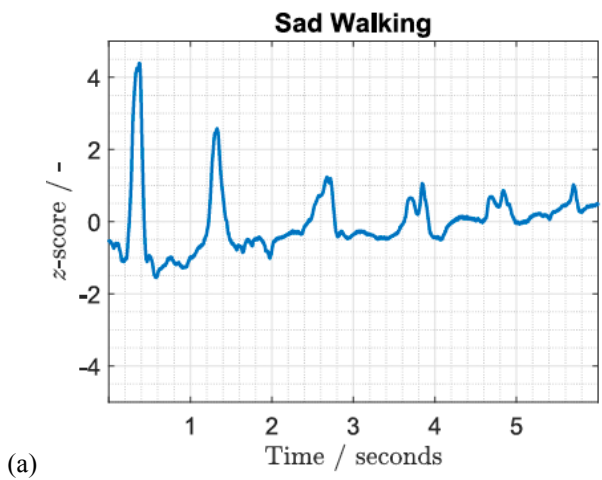

(a)

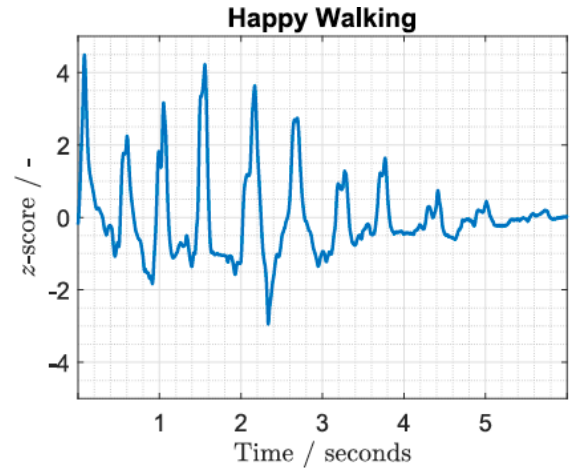

(b)

**Figure 4.** Sample signals of one person walking sadly (a) and happily towards a plant (b)

## 4. PREDICTING INDIVIDUAL PEOPLE THROUGH THEIR GAIT

Similarly, electrostatic discharge of 6 people was recorded, identifying the correct person based on her/his gait. Two electrical signals gathered from two different persons walking by a plant are shown in Figure 5. The amplitude is increasing while walking towards and decreasing when walking away from the plant in both signals. While the frequency of steps seems to be identical, the way the foot rises and moves is different. The shapes around the area of the peaks change accordingly. The envelope of our gathered signals is comparable to the simulations and measurements of other studies in electric field sensing (Chen et al. 2012). Using a random forest classifier with 100 trees, and applying it on 212 samples classified into six classes, one per person, 66% accuracy was achieved (Cohens κ=0.51, AUROC=0.876).

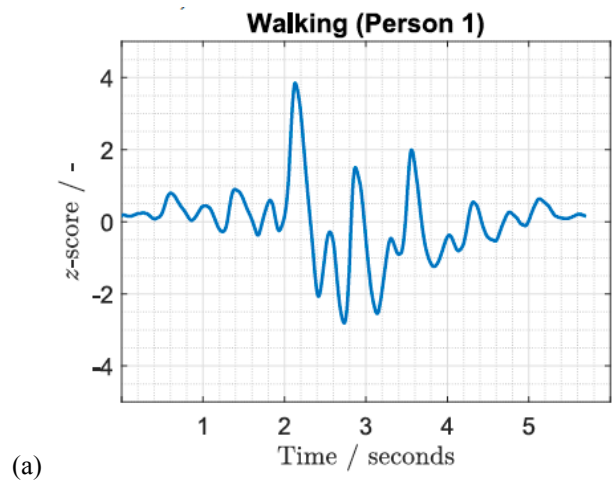

(a)

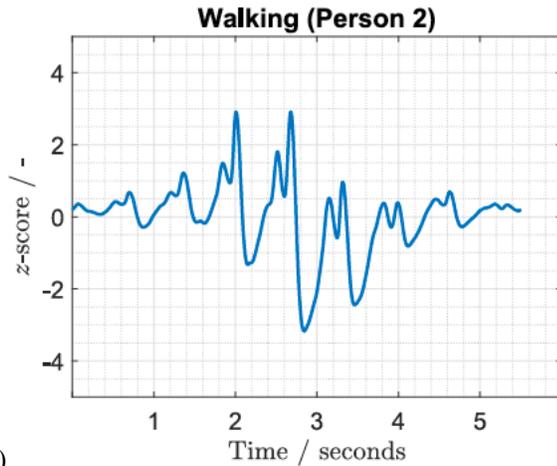

(b)

**Figure 5.** Sample signals of two different people walking by Mimosa Pudica

In figure 6, we compared the random forest classifier to a baseline 0R-classifier, which always predicts the class with the most samples. When only two classes are to be predicted, the accuracy of the random forest classifier reaches 88% Accuracy comparable to the classification of "happy" and "sad" walking. With more than four people, the accuracy of the random forest classifier stays constant at 66%, whereas the accuracy of the 0R-classifier falls because of the additional diversification. This trend shows that our approach could be used in crowded areas or in public events.

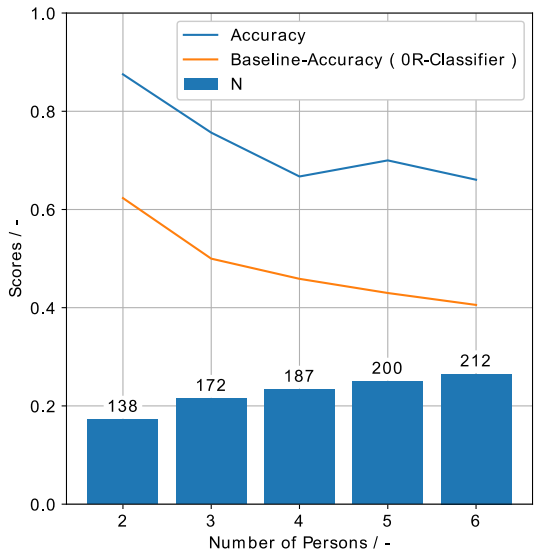

**Figure 6.** Accuracy of random forest classifier in comparison with baseline classifier over the number of persons

Besides the 20 MFCC features, the plant type and the location were included in the prediction, to investigate their influence. Random forests can be used for feature selection. When the algorithm is fit, features that are not useful will not be used to split the data (Breiman 2001). For calculating the feature importance, we used the Mean Decrease in Impurity (MDI). Proportionally to the number of samples the tree splits, the sum over the splits across all trees is calculated. As mentioned above, besides Mimosa Pudica, different plants were used. The measurements were taken in Boston at two different locations, and as well as in Switzerland. As figure 7 illustrates, location and plant type were not essential predictors; the MFCC features from the electrostatic discharge signal were far more predictive. The feature importance indicates that the type of floor, the type of plant, and the location do not play a significant role in identifying a person. The lower order coefficients of the MFCCs and thus the lower frequencies of the electrical signal contain most of the information about the individual human gait characteristics.

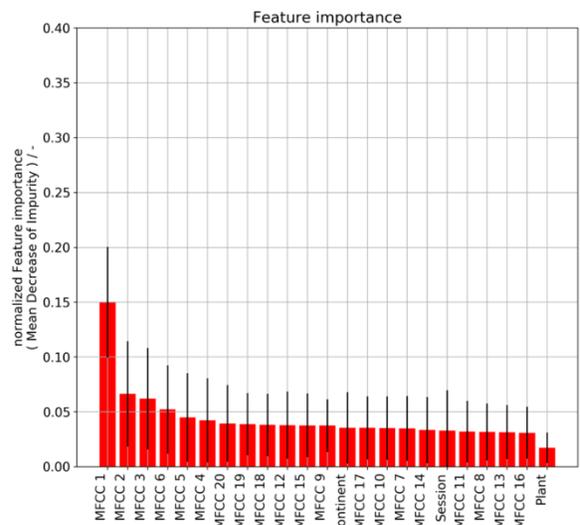

**Figure 7.** Feature importance for predicting different people

## 5. MEASURING LEG SHAKING WITH PLANTS

Leg shaking is often viewed as a negative behavior and is a common human stereotype that many people exhibit while sitting down. A study (Niehaus et al. 2000) found that over half of the study's population (N=750) experienced some form of leg shaking. Leg shaking was reported to disrupt normal function and carry a negative social stigma while also being linked to medical conditions such as Attention Deficit Hyperactivity Disorder (ADHD). However, subjects without these conditions may also experience leg shaking due to anxiousness or agitation. Additionally, significant relations between the Big Five personality traits and leg shaking have been found. (Oshio 2018)

To see how well our method works with leg shaking, we let individuals sit on a chair and observed their leg shaking besides the plant. In figure 8, the measured electrical signal is shown over time. As soon as the individual starts shaking their leg, a distinct pattern with a frequency of 5-6 Hz is visible.

(Xia et al. 2018) studied leg shaking prediction using accelerometers. They found that the power peak of the signal is around 6 Hz. Using the ratio of frequencies and feeding it into a Bayes' classifier, they were able to get an accuracy of greater than 90%. As shown in figure 8, we were able to pick up the same spectrum of signals using our method.

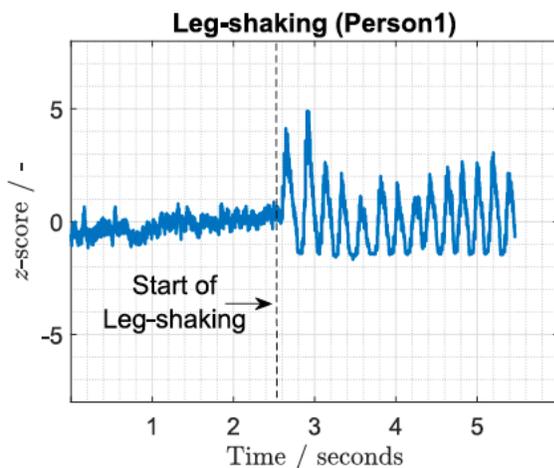

**Figure 8**. Sample signals of one individual shaking their legs besides the plant

Our system would have applicability for instance for patients without medical conditions like Restless Leg Syndrome, so they can voluntarily halt their leg shaking once aware of the motion. The individual can be notified in real time once our setup detects leg shaking.

## 6. CONCLUSION AND FUTURE WORK

In this paper, we investigated how plants and humans can be connected through electric field sensing. Simple data acquisition systems and electrodes on plants enable us to pick up the electrostatic discharge of humans that are passing by. These signals contain information on gait characteristics that can be extracted using features like MFCC. With a 10-fold cross-validated random forest classifier, it was possible to predict one out of six humans that are walking by the plant with 66% accuracy. Dysphoric mood observable in gait-patterns of these humans could be predicted with an accuracy of 88%. Similar to gait-characteristics, the behavior of sitting individuals' leg-shaking can be observed using our system. Further analysis revealed that low-frequency MFCCs are the most predictive feature for walking patterns. The type of plant used, and the location where the experiment was conducted were not significant for the prediction.

Applications for this plant-based sensing system could be used in large scale office environments or public events. Information about the happiness and other emotions of customers and employees could be used to improve office climate and performance. Smartwatch-based collaboration measurement tools such as the Happimeter (Roessler & Gloor 2020) could be upgraded to enhance mood prediction using plants as emotion sensors.